\documentstyle[11pt]{article}

\evensidemargin=\oddsidemargin
\begin{document}
\begin{flushright}
TUIMP-TH-95/62\\

February 1995\\
\end{flushright}
\vspace{.8cm}
\begin{center}
{\large \bf RADIATIVE CORRECTIONS TO THE $ Zb {\bar b} $ and
$ Z \tau^+ \tau^- $ VERTICES IN A REALISTIC ONE-FAMILY EXTENDED
TECHNICOLOR MODEL  }\\
\vspace{1cm}
 {\bf Chong-Xing Yue$^{a,b} $, Yu-Ping Kuang$^{a,c} $,    \\
 Gong-Ru Lu $^{a,b} $, and Ling-De Wan $^b $ } \\
{\small a. China Center of Advanced Science and Technology
 (World Laboratory), P. O. Box 8730, Beijing 100080, China}\\
{\small b.  Physics Department, Henan Normal University, Xin Xiang, }\\
{\small Henan 453002,  P. R. China }\footnote{Mailing address}\\
{\small c. Institute of Modern Physics,Tsinghua University, Beijing, }\\
{\small 100084, P. R. China. } \\
\end{center}
\vspace{1.8cm}
\centerline{\bf Abstract}
\vspace{0.2cm}
\begin{sf}

In a realistic effective one-family extended technicolor(ETC) model without
exact custodial symmetry, we calculate the one-loop corrections to the
$~Zb {\bar b}~$ and $~Z \tau^+\tau^-~$ vertices from the sideways and
diagonal ETC bosons exchange. The result shows that both the $~Z\rightarrow
b{\bar b}~$ partial width $ \Gamma_b $ (and branching ratio $~R_b~$)
and the $~\tau~$ polarization
asymmetry parameter $ A_{\tau} $ are enhanced by the corrections and are in
agreement with the present experimental data.\\
\vspace{0.4cm}
\noindent
PACS numbers: 12.60Nz, 13.38Dg
\end{sf}

\newpage
\begin{sf}
\section{Introduction }
\indent

  Technicolor(TC) models$^{[1]} $ can give contributions to electroweak
observables. The precision measurements of electroweak  observables give
strong constraints on TC models. The general approach to confront
TC models with precision measurements is based on the parametrization of the
vacuum polarizations and vertex corrections. Specifically, the constraints
are characterized by the three independent parameters S, T, and U$^{[2]}$
( or $ \epsilon_{1, 2, 3} )^{[3]}$ in the oblique corrections and the
corrections to the $~Zb{\bar b}~$ $^{[4,5]}$ and $~Z\tau^+\tau^-~$$^{[6]}$
vertices. It has been shown${[2]}$ that one-family TC models with exact
custodial symmetry seem to be already excluded by the experimental value of
the parameter S. However, as is shown in Ref.${[7]}$, this is not the case for
TC models without exact custodial symmetry. A realistic model is proposed for
a one-family TC model in Ref.$[7] $. In this model, isospin is a good
approximate symmetry only for techniquarks but is broken for technileptons.
Such an isospin splitting can contribute to the parameter S with negative sign
without making a large contribution to the parameter T. Recently, it has been
shown that this kind of ETC model$^{[8]}$ can further explain the puzzling
feature of the quark-lepton mass spectrum.

  The "high-energy" (heavy TC particle) contributions to the $ Zb {\bar b} $
vertex from "integrating out" technifermions have been given in Refs.$[4]$ and
$[5]$. These contributions decrease the $ Z b {\bar b} $ branching ratio
$ R_b $ and are large , even in walking technicolor models$^{[9]} $. There are
also "low-energy" (light TC particle) contributions to the $ Zb {\bar b} $
vertex coming from the Pseudo Goldstone bosons (PGB's). For one-family TC
model with exact custodial symmetry, the effects have been found to decrease
$ R_b $ by a few percent$^{[10]}$. The current LEP data $^{[11]}$ is likely to
exclude TC models if the top mass is $100 Gev $ or greater. Noncommuting ETC
models $^{[12]}$ have been proposed to give a positive
correction to the ratio $ R_b $. But it is important to consider at the
same time the $ S $ parameter. More recently it has been found$^{[13]}$ that
an extra gauge boson existing in certain types of dynamical electroweak
symmetry breaking models can also give positive corrections to $ R_b $
and the
$\tau $ polarization asymmetry parameter $ A_{\tau} $.

  In this paper we reconsider and carefullly calculate the non-oblique
corrections to the $ Z b {\bar b} $ and  $ Z \tau^+\tau^- $
vertices from ETC dynamics in a one-family ETC model without exact custodial
symmetry. The purpose of this paper is to study the phenomenological aspect of
the ETC theory without concerning explicitly the ETC model. So we simply
adopt the realistic toy model description of Ref.[5] in our calculation, in
which the right-handed top quark and bottom quark are put in the same
representation and their mass difference are described by assigning different
effective ETC coupling constants to them. We find that the sideways ETC boson
exchange decreases the width $ \Gamma_b = \Gamma(Z \rightarrow b {\bar b}) $
and $ R_b =\frac{ \Gamma_b}{\Gamma_h} $, while the diagonal ETC boson exchange
increases $ \Gamma_b $ and $ R_b $ which is contrary to the result in Ref.
$[5]$. In this kind of model $^{[7]}$, the decay constant of the technipion
in the technilepton sector is significantly smaller than that in the
techniquark sector. So the corrections to the $ Z \tau^+\tau^-$ vertex is
much smaller than the corrections to the $ Z b {\bar b} $ vertex.
However, we find that the correction to the $ \tau $ asymmetry parameter
$A_{\tau} $ is positively enhanced and is in agreement with the present
experimental data.

  This paper is organized as follows. In Sec.II we discuss the masses of
quarks and leptons in a realistic effective one-family ETC model. The
calculations of the corrections to the $ Z b {\bar b} $ vertex and the
$ \tau $ polarization asymmetry parameter $ A_{\tau} $ from the ETC boson
exchanges will be presented in Sec.III and Sec.IV, respectively. Sec.V is a
concluding remark.

\section{The Masses of Quarks and Leptons}
\indent

  To generate the masses of quarks and leptons, the ETC gauge group
$ SU(N_{TC}+1) $ is assumed to hierarchically break down to the TC gauge group
$ SU(N_{TC} ) $. In this process many ETC gauge bosons become massive. Some of
them mediating ordinary fermions with technifermions are called sideways
bosons and some of them mediating the same kind of fermions are called
"diagonal" bosons. The sideways ETC bosons must exist in any realistic ETC
models to generate the masses of quarks and leptons, while the existence of
diagonal ETC bosons is model-dependent. In the present model, there are
both knds of ETC bosons.

 The approximate global chiral symmetry of the present one family model
is $ G = SU(6)_L \times SU(6)_R \times U(1)_{2R} \times U(1)_{8L} \times
U(1)_{8R} \times U(1)_V $$^{[7]}$. The mass spectrum of technifermions is

\begin{eqnarray}
(a), M_U \approx M_D , \hspace{2cm}   (b) M_N < M_E < M_U   \nonumber
\end{eqnarray}
These technifermions can be assigned to the following representations of
$ SU(3)_C \times SU(2)_L \times U(1)_Y \times SU(N_{TC} ) $
\begin{eqnarray}
 Q_L = (U, D)_L = (3, 2, \frac{Y_{Lq} }{2}, N_{TC} ),
 U_R = (3, 1, \frac{Y_{Lq}}{2}+\frac{1}{2}, N_{TC})       \nonumber
\end{eqnarray}
\begin{eqnarray}
 D_R = (3, 1, \frac{Y_{Lq}}{2}- \frac{1}{2}, N_{TC}) ,
 L_L = (N, E)_L = (1, 2, \frac{Y_{Ll}}{2}, N_{TC})   \nonumber
\end{eqnarray}
\begin{eqnarray}
 E_R = (1, 1, \frac{Y_{Ll}}{2}- \frac{1}{2}, N_{TC}) ,
 N_R = (1, 1, \frac{Y_{Ll}}{2} + \frac{1}{2}, N_{TC}),   \nonumber
\end{eqnarray}
where $ Y_{Lq} = \frac{1}{3}$ and $ Y_{Ll} = -1 $ are the hypercharges of
the left-handed techniquark and technilepton, respectively.

 The lagrangian in the model describing the sideways ETC gauge interaction
between the third family fermion and
 technifermion is$^{[5],[6]}$
\begin{eqnarray}                                  
L = g_E(\xi_t \overline {Q_L} W_E^{\nu} \gamma_{\nu} q_L +\xi_{Rt}
\overline {U_R}W_E^{\nu} \gamma_{\nu} t_R
+ \xi_b \overline{D_R} W_E^{\nu} \gamma_{\nu} b_R + h.c.    \nonumber
\end{eqnarray}
\begin{eqnarray}
+ \xi_{\tau} \overline{L_L} W_E^{\nu} \gamma_{\nu} l_L + \xi_{\nu}
\overline {N_R} W_E^{\nu} \gamma_{\nu}{\nu}_R+ \xi_{R \tau} \overline{E_R}
W_E^{\nu} \gamma_{\nu}{\tau}_R + h.c.) ,
\end{eqnarray}
where $ q_L= (t, b)_L, t_R $ and $ b_R $ represent the third family of
quarks,  $ l_L = (\nu, \tau )_L, \nu_R $ and $ \tau_R $ represent the third
family of leptons, $W_E$ is the sideways ETC boson coupling to the third
family ordinary fermions and technifermions with the coupling constant $ g_E $,
 and $ \xi_i $ is the coefficient of the left-handed or right-handed couplings
$^{[6]}$. The summation over the color index is implied in (1).

  Using the rules of naive dimensional analysis $^{[14]}$, the ordinary
fermion mass can be written as $ m_i = \frac{g_E^2}{m_E^2}(4 \pi f_i^3) $.
{}From (1) the masses of ordinary fermions are given as
\begin{eqnarray}                                 
m_t = \xi_t \xi_{Rt} \frac{g_E^2}{m_S^2}<\overline {U} U> =
 \xi_t \xi_{Rt} \frac{g_E^2}{m_S^2}4 \pi f_Q^3
\end{eqnarray}
\begin{eqnarray}                                  
m_b =  \xi_t \xi_b \frac{g_E^2}{m_S^2}<\overline {D}D > =
\xi_t \xi_b \frac{g_E^2}{m_S^2}4\pi f_Q^3
\end{eqnarray}
\begin{eqnarray}                                 
m_{\nu} =  \xi_{\tau} \xi_{\nu} \frac{g_E^2}{m_S^2}<\overline {N}N> =
 \xi_{\tau} \xi_{\nu}\frac{g_E^2}{m_S^2} 4 \pi f_N^3
\end{eqnarray}
\begin{eqnarray}                                  
m_{\tau} = \xi_{\tau} \xi_{R\tau} \frac{g_E^2}{m_S^2}<\overline{E}E > =
\xi_{\tau} \xi_{R\tau}\frac{g_E^2}{m_S^2}4 \pi f_E^3 ,
\end{eqnarray}
where $ m_S $ is the mass of the sideways ETC boson, and $ f_Q, f_E, f_N $
are
the decay constants of technipions composed of techniquarks, technielectron
and technineutrino, respectively.

 In one-family TC models with custodial symmetry, the technipion decay
constant is $ 4 F_{\pi}^2 = (250Gev)^2 $. In models without custodial symmetry,
the decay constant in the technileption sector is different from that in the
techniquark sector. They satisfy the following constraint:$^{[7]}$
\begin{eqnarray}                                 
N_C f_Q^2 + \frac{1}{2}f_N^2 + \frac{1}{2}f_E^2 \approx (250Gev)^2 ,
\end{eqnarray}
where $ N_C = 3 $ is the number of colors in QCD. In the realistic one-family
TC model$^{[7]}$, the decay constants are taken to be $~N_C f_Q^2 \gg
\frac{1}{2}(f_N^2 + f_E^2)~$ in order to keep the electroweak parameter S
small or even negative without violating the experimental bound on the
parameter T.

 For given quark and lepton masses, we see from (2)-(5) that the $ \xi_i'$s
should have the relations
\begin{eqnarray}                                 
 \xi_t =  \xi_{Rt}^{-1}, \hspace{2cm}   \xi_b = \xi_t^{-1} \frac{m_b}{m_t}
\end{eqnarray}
\begin{eqnarray}                                 
 \xi_{\tau} =  \xi_{R\tau}^{-1} , \hspace{2cm}   \xi_{\nu} = \xi_{\tau}^{-1}
\frac{m_{\nu}}{m_{\tau}} .
\end{eqnarray}
If we take $ m_t = 175Gev, m_b = 4.8Gev$ , and $m_{\nu} \approx 0$, we have
$~\xi_b \approx 0.028 \xi_t^{-1}$ and $ \xi_{\nu} \approx 0 $.

\section{ Corrections to the $Zb{\bar b}$ Vertex}
\indent

  The corrections to the $ Z b {\bar b} $ vertex can come from sideways
ETC boson exchange and diagonal ETC boson exchange. From (1) , the effective
four-fermion operators from sideways ETC boson exchange are
\begin{eqnarray}                                   
- \frac{g_E^2}{m^2_S}[ \xi_t^2(\overline{Q_L} \gamma^{\nu} q_L)(\overline q_L
\gamma_{\nu} Q_L) + \xi_b^2(\overline{D_R}
\gamma^{\nu}b_R)(\overline{ b_R} \gamma_{\nu} D_R)]
\end{eqnarray}
We can further make Fierz reordering to express the four-fermion operators
in terms of quark-quark currents and techniquark-techniquark currents. Note
that only color-singlet currents can couple to the $Z$-boson, and are thus
relevant to the $~Zb{\bar b}~$ vertex. The piece composed of color-octet
currents is irrelevant to the present study. After Fierz reordering, the
relevant four-fermion operators are
\begin{eqnarray}                                   
- \frac{g_E^2}{2 m^2_S}\frac{1}{N_C}[\xi^2_t(\overline{Q_L} \gamma^{\nu}
\tau^a Q_L)
(\overline {q_L}\gamma_{\nu} \tau^a q_L) + \xi^2_b(\overline{D_R}
\gamma^{\nu} D_R)(\overline b_R \gamma_{\nu} b_R)]  ,
\end{eqnarray}
where color and technicolor summation is implied and $ \tau^a (a = 1, 2, 3 )$
are weak isospin Pauli matrices.

  Adopting an effective chiral lagrangian description which is appropriate
below the technicolor chiral symmetry  breaking scale, the technifermion
current may be replaced by the corresponding sigma model current$^{[15]}$.
\begin{eqnarray}                                  
(\overline{ Q_L} \gamma_{\nu} \tau^a Q_L ) =  \frac{N_C f_Q^2}{2}
Tr (\Sigma^+ \tau^a i D_{\nu L} \Sigma )
\end{eqnarray}
\begin{eqnarray}                                  
(\overline{ D_R} \gamma_{\nu} D_{R} ) =  \frac{N_C f_Q^2}{2}
Tr(\Sigma i D_{\nu R} \Sigma^+ )  ,
\end{eqnarray}
where $ \Sigma = exp(\frac{2i\phi}{ f_Q } ) $ transforms as
$ \Sigma \rightarrow L \Sigma R^+ $ under $ SU(2)_L\times SU(2)_R $,
$ \phi $ is the Nambu - Goldstone boson field, and the covariant derivatives
$ D_{\nu L}$ and $ D_{\nu R } $ are
\begin{eqnarray}                                   
D_{\nu L}\Sigma = \partial_{\nu} \Sigma + i \frac{ e}{ \sqrt 2 S_{\theta} }
(W^+_{\nu} \tau^+ + W^-_{\nu} \tau^- )\Sigma    \nonumber  \\
+ i \frac{e}{ S_{\theta } C_{\theta} } Z_{\nu} (\frac{1}{2} \tau_3
\Sigma -
S^2_{\theta}[ Q, \Sigma ] ) + ieA_{\nu} [Q, \Sigma]   ,
\end{eqnarray}
\begin{eqnarray}                                   
D_{\nu R}\Sigma = \partial_{\nu} \Sigma - i \frac{e}{ S_{\theta}
C_{\theta}}
Z_{\nu} (\frac{1}{2} \tau_3 \Sigma + C^2_{\theta}\{ Q, \Sigma \} ) +
ieA_{\nu}
[Q, \Sigma]  .
\end{eqnarray}
In the unitary gauge $ \Sigma = 1 $, we can give the terms which are
relevant to the $ Zb {\bar b} $ vertex from the operator $ (10) $,
\begin{eqnarray}                                     
\frac{g^2_E }{ 2 m^2_S }\frac{f_Q^2 e}{S_{\theta }C_{\theta}}
[ \xi^2_t \overline{q_L} Z \frac{\tau_3}{ 2 } q_L -  \xi^2_b
\overline{b_R} Z \frac{\tau_3}{2} b_R ] ,
\end{eqnarray}
where $ S_{\theta} = \sin \theta, C_{\theta} = \cos \theta $ with
$\theta $ being the Weinberg angle. These yield the corrections to
the tree-level  standard model $ Zb {\bar b}$ couplings
\begin{eqnarray}                                    
g_L^b = \frac{ e}{ S_{\theta }C_{\theta}}(- \frac{1}{ 2} +\frac{1}{ 3}
S^2_{\theta} ) ,   \hspace{2cm}
g_R^b = \frac{ e}{ S_{\theta }C_{\theta}}( \frac{1}{3} S^2_{\theta} )  .
\end{eqnarray}
The corrections are
\begin{eqnarray}                                     
\delta g_{LS}^b = \frac{ \xi^2_t}{4 } \frac{g^2_E f_Q^2}{m^2_S }
\frac{e}{ S_{\theta }C_{\theta}} =\frac{ \xi^2_t}{4} \frac{m_t}
{4 \pi f_Q }
\frac{ e}{ S_{\theta }C_{\theta}} ,
\end{eqnarray}
\begin{eqnarray}                                     
\delta g_{RS}^b =-\frac { \xi^2_b}{ 4 } \frac{g^2_E f_Q^2}{m^2_S }
\frac{ e}{S_{\theta }C_{\theta}} = - \frac{ \xi^2_b}{4} \frac{m_t}
{4 \pi f_Q }
\frac{ e}{ S_{\theta }C_{\theta}}  .
\end{eqnarray}
Since the tree level $ Z b_L \overline{b_L} $ coupling $ g_L^b $
is negative and $ Z b_R \overline{b_R} $ coupling $ g_R^b $ is positive, we
see that the sideways ETC boson exchange corrections decrease the width
$ \Gamma_b $ relative to the standard model prediction. This is
disfavored by the recent precision electroweak measurements.

 Aprt from the sideways boson contributions, there are also diagonal boson
contributions in the present realistic model$^{[5]}$. For a given technicolor
number $ N_{TC} $, we can obtain the diagonal coupling of technifermions by
multiplying the factor $ \frac {1}{\sqrt {N_{TC}(N_{TC}+1)}}$ to
their sideways coupling, and that of ordinary fermions by multiplying the
factor $ - \sqrt{\frac {N_{TC}}{N_{TC}+1}} $ to their sideways coupling. These
factors come from the normalization of the tranceless diagonal generator. The
diagonal boson exchange gives rise to the following four-fermion operators
\begin{eqnarray}                             
 (\frac{m_S}{m_D})^2 \frac{g_E^2}{2m_S^2} \frac{1} {N_{TC} +1}
[(\overline{U_R}\gamma^{\nu} U_R)(\overline {q_L}
\gamma_{\nu}q_L) +
 \xi_t \xi_b (\overline{D_R}\gamma^{\nu}D_R)
(\overline{ q_L} \gamma_{\nu} q_L)    \nonumber
\end{eqnarray}
\begin{eqnarray}
+ \xi_t^{-1} \xi_b(\overline {U_R}\gamma^{\nu} U_R)(\overline{ b_R}
\gamma_{\nu}b_R)] .
\end{eqnarray}
Similar to (17) and (18), we obtain the following corrections to $g_L^b$ and
$g_R^b$ from the diagonal ETC boson exchange
\begin{eqnarray}                             
\delta g_{LD}^b = -\frac{1}{4} \frac{m_t}{4 \pi f_Q }
\frac{ e}{ S_{\theta }
C_{\theta}}(\frac{m_S}{m_D})^2 \frac{N_C}{N_{TC}+ 1} \xi_t( \xi_t^{-1}
+  \xi_b)  ,
\end{eqnarray}
\begin{eqnarray}                            
\delta g_{RD}^b = - \frac{1}{4} \frac{m_t}{4 \pi f_Q }
\frac{ e}{ S_{\theta }
C_{\theta}}(\frac{m_S}{m_D})^2 \frac{N_C}{N_{TC}+ 1} \xi_t^{-1} \xi_b  .
\end{eqnarray}
We see that the diagonal ETC boson exchange gives negative corrections to both
$ g_L^b $ and  $ g_R^b $, and its total effect is to increase $\Gamma _b $
and $ R_b$ contrary to that from the sideways ETC boson exchange. This result
differs from that in Ref.$[5]$ by a minus sign (ref$[16]$ gives the similar
conclusion). Note that $\delta g^b_{LD}$ and $\delta g^b_{RD}$ decrease
as $N_{TC}$ increases. Furthermore, we see from (18), (21), and (7) that
$\delta g^b_{RS}$ and $\delta g^b_{RD}$ are suppressed by $(\frac
{m_b}{m_t})^2 $ and $( \frac {m_b}{m_t}) $, respectively. Therefore,
$\delta g_{RD}$ is more important than $\delta g_{RS}$, and can not be
ignored.

  Summing up the corrections to the $ Z b {\bar b} $ vertex from the
sideways and diagonal ETC boson exchange , we obtain the total
correction
\begin{eqnarray}                               
\delta g_{LE}^b = -\frac{1}{4} \frac{m_t}{4 \pi f_Q }
\frac{ e}{ S_{\theta }
C_{\theta}}[(\frac{m_S}{m_D})^2 \frac{N_C}{N_{TC}+ 1} \xi_t(
\xi_t^{-1} +
\xi_b) - \xi^2_t] ,
\end{eqnarray}
\begin{eqnarray}                                
\delta g_{RE}^b \approx - \frac{1}{4} \frac{m_t}{4 \pi f_Q } \frac{ e}
{ S_{\theta }C_{\theta}}(\frac{m_S}{m_D})^2 \frac{N_C}{N_{TC}+ 1}
\xi_t^{-1}
 \xi_b  .
\end{eqnarray}
We see that that magnitudes of these two corrections are comparable. For
large
$N_{TC}$, the sideways ETC boson exchange contribution to $\delta g^b_{LE}$
may dominate. However, $m_S/m_D $ is a model-dependent parameter. Therefore,
it is possible to have a realistic model leading to positive corrections to
$\delta g^b_{LE}$ and $\delta g^b_{RE}$.

 Having (22) and (23), we obtain the following total corrections to $\Gamma_b$
and $R_b$
\begin{eqnarray}                                   
(\frac{\delta \Gamma}{ \Gamma_b})_E = \frac{ 2(g_L^b \delta g_{LE}^b
+ g_R^b \delta g_{RE}^b )}{ g^{b2}_L + g^{b2}_R }    \nonumber
\end{eqnarray}
\begin{eqnarray}
\approx +1.26 \% (\frac{m_t}{175Gev} )  ,
\end{eqnarray}
\begin{eqnarray}                                    
\delta R_{bE} = \delta(\frac{ \Gamma}{ \Gamma_b})_E = (\frac{\delta \Gamma}
{ \Gamma_b})_E (\frac{ \Gamma_b}{ \Gamma_h})
(1 - \frac{\Gamma_b}{ \Gamma_h})   \nonumber
\end{eqnarray}
\begin{eqnarray}
\approx +0.99 \% (\frac{m_t}{175Gev} )R_b .
\end{eqnarray}
In (24) and (25), we have take $\xi_t \approx \frac{1}{\sqrt 2},
\xi_b \approx 0.028 \xi_t^{-1}, N_{TC}= 4 , f_Q \approx 143Gev$
and $m_S \approx m_D $.

 For the ratio $R_b$, all oblique corrections and the leading QCD corrections
cancel. $ \delta R_{bE} $ is the pure nonolique correction to the
the $Zb{\bar b}$ vertex. The recent measurement of $ R_b $ at the CERN
$ e^+e^- $ collider LEP$^{[11]}$ gives $ R_b = 0.2202 \pm0.0020 $ which is
larger than the standard model predicted value $R_b^{SM} = 0.2157
 \pm 0.0004$ by more than two standard deviations. So the above result in (25)
is particularly interesting.

So far we have focused on the " high energy'' technicolor corrections to
the $ Z b {\bar b} $ vertex from ETC boson exchange, there are still " low
energy '' corrections coming from the PGB's$^{[10]}$. However, the masses of
the PGB's could be significantly enhanced in the realistic one-family ETC
model, so that the PGB's correction to $R_b$ could be significantly reduced
and could thus be ignored.

\section{Corrections to the $ Z \tau^+\tau^- $ Vertex}
\indent

 Now we calculate the corrections to the $ Z \tau^+\tau^-$ vertex from the
sideways and diagonal ETC boson exchanges. We particularly estimate the effect
of these corrections on the $ \tau $ polarization asymmetry parameter
$ A_{\tau} = \frac {(g_L^{\tau})^2 - (g_R^{\tau})^2 }{(g_L^{\tau})^2 +
(g_R^{\tau})^2} $ whose experimental value serves as the best constraint on
the $Z\tau^+\tau^-$ vertex.

  From (1) we can write down the four-fermion operators contributing to the
$Z \tau^+\tau^-$ vertex from the sideways ETC boson exchange. After Fierz
transformation it becomes
\begin{eqnarray}                                
- \frac{g_E^2}{2m^2_S}[\xi_{\tau}^2(\overline{L_L} \tau^a\gamma^{\nu}
 L_L)(\overline l_L \gamma_{\nu} \tau^a l_L) + \xi_{\tau}^{-2}
(\overline{E_R} \gamma^{\nu} E_R)(\overline{\tau_R} \gamma_{\nu}
 \tau_R)] .
\end{eqnarray}
Similar to the derivation of (19), we can also obtain the four-fermion
operators from the diagonal ETC boson exchange
\begin{eqnarray}                                
  (\frac{m_S}{m_D})^2 \frac{g_E^2}{2m_S^2} \frac{1} {N_{TC} +1}
\{(\overline{E_R} \gamma^{\nu} E_R)(\overline {l_L} \gamma_{\nu}l_L)
+ \xi_{\tau}[ \xi_t^{-1}(\overline{U_R}\gamma^{\nu} U_R)
+ \xi_b( \overline{D_R}\gamma^{\nu} D_R)](\overline {l_L}
\gamma_{\nu} l_L)     \nonumber
\end{eqnarray}
\begin{eqnarray}
+ \xi_{\tau}^{-1}[ \xi_t^{-1}(\overline{U_R}\gamma^{\nu} U_R)
+ \xi_b (\overline{D_R}\gamma^{\nu} D_R)](\overline {\tau_R}
\gamma_{\nu} \tau_R)\} .
\end{eqnarray}

 By means of the effective lagrangian approach we can obtain the corrections
to the tree-level vertex of $ Z \tau^+\tau^- $ coupling
\begin{eqnarray}                                 
g_L^{\tau} = \frac{ e}{ S_{\theta }C_{\theta}}(- \frac{1}{ 2} +
S^2_{\theta} ) ,   \hspace{2cm}
g_R^{\tau} = (\frac{ e}{ S_{\theta }C_{\theta}}) S^2_{\theta} ,
\end{eqnarray}
which are
\begin{eqnarray}                                 
\delta g_{LE}^{\tau} = -\frac{1}{4} \frac{m_t}{4 \pi} \frac{1}{f_Q }
 \frac{ e}{ S_{\theta }C_{\theta}}[-\frac{f_E^2}{f_Q^2}\xi_{\tau}^2
+(\frac{m_S}{m_D})^2 \frac{N_C}{N_{TC}+ 1}\xi_{\tau}
 (  \xi_t^{-1}+ \xi_b +\frac{f_E^2}{N_Cf_Q^2} \xi_{\tau}^{-1})] ,
\end{eqnarray}
\begin{eqnarray}                                 
\delta g_{RE}^{\tau} \approx - \frac{1}{4} \frac{m_t}{4\pi}
\frac{1}{f_Q }\frac{ e}{S_{\theta }C_{\theta}}
[-\frac{f_E^2}{f_Q^2}\xi_{\tau}^{-2}+(\frac{m_S}{m_D})^2
\frac{N_C}{N_{TC}+ 1}\xi_{\tau}^{-1}( \xi_t^{-1}+ \xi_b )] .
\end{eqnarray}
Considering the relation $f_Q^2 \gg f_E^2 $ the above results can be
approximately written as
\begin{eqnarray}                                  
\delta g_{LE}^{\tau} = - \frac{1}{4} \frac{m_t}{4\pi}\frac{1}{f_Q }
\frac{ e}{S_{\theta }C_{\theta}}(\frac{m_S}{m_D})^2
\frac{N_C}{N_{TC}+ 1}\xi_{\tau}( \xi_t^{-1}+ \xi_b ) ,
\end{eqnarray}
\begin{eqnarray}                                  
\delta g_{RE}^{\tau} \approx - \frac{1}{4} \frac{m_t}{4\pi}
\frac{1}{f_Q }\frac{ e}{S_{\theta }C_{\theta}}(\frac{m_S}{m_D})^2
\frac{N_C}{N_{TC}+ 1}\xi_{\tau}^{-1}( \xi_t^{-1}+ \xi_b ) .
\end{eqnarray}

The relative correction $\frac{\delta A_\tau}{A_\tau}$ can be expressed
in terms of $\delta g_L^\tau$ and $\delta g_R^\tau$ as
\begin{eqnarray}                                     
\frac{\delta A_{\tau}}{A_{\tau}} = \frac {4 (g_L^{\tau})^2 (g_R^{\tau})^2 }
 {(g_L^{\tau})^4 - (g_R^{\tau})^4} (\frac {\delta g_L^{\tau}}{g_L^{\tau}}
 - \frac {\delta g_R^{\tau}}{g_R^{\tau}})   \nonumber
\end{eqnarray}
\begin{eqnarray}
 = - (\frac{ e}{S_{\theta }C_{\theta}})^{-1} (24.04 \delta g_L^{\tau}
 + 27.99\delta g_R^{\tau}) .
\end{eqnarray}
{}From (32) and (33) we see that $(\frac{\delta A_{\tau}}{A_{\tau}}) >0 $.
The recent measurement of the parameter $A_{\tau} $ is
$ \frac{\delta A_{\tau}}{A_{\tau}} = 0.31\pm 0.13^{[13]} $. So the realistic
one-family ETC model without custodial symmetry is also consistent with the
present experiment of $A_\tau$.

The PGB's correction to the $ Z \tau^+\tau^- $ vertex is suppressed by the
square of the small $ \tau $ mass and can thus be ignored$^{[16]}$.
However, there is another large correction from the isospin breaking effect in
the tecnileption sector coming from the vector mesons composed of
technileptions. The corrections to the $ Z \tau^+\tau^- $ vertex coming from
neutral techni-vector-mesons can not be simply ignored. In Ref.$[6]$, such
correction to the coupling constant $ g_L^{\tau} $ has been computed and the
difference between the corrections to the $ Z \tau_L^+ \tau_L^- $ and the
$ W \tau \nu $ vertices has been given. We shall give a detailed examination
of such correction to the coupling constant $ g_R^{\tau} $ in our future work.

\section{Conclusions}
\indent

 We have calculated the corrections to the  $ Z b {\bar b} $ vertex and the
$ Z \tau^+\tau^- $ vertex from ETC boson exchanges in a realistic one-family
ETC model with diagonal ETC boson. We have estimated the effect on the
$Zb{\bar b}$ couplings from both the sideways ETC boson exchange and the
diagonal ETC boson exchange. We find that the sideways ETC boson exchange
decreases the width $\Gamma_b $, while the diagonal ETC boson exchange tends
to increase it. The two corrections are comparable and the total correction
to the $Zb{\bar b}$ vertex gives a positive correction to $R_b$ which is
consistent with the present LEP experiment. We have also calculated the
corrections to the tree-level vertex of $Z\tau^+\tau^-$ couplings $g_L^{\tau}$
and $g_R^{\tau}$ from the ETC boson exchanges. We find that the correction to
$g_R^{\tau}$ is not smaller than that to $g_L^{\tau}$ and can not be ignored.
We then give an estimate of $ \frac{\delta A_{\tau}}{A_{\tau}} $ from the
obtained $\delta g_L^\tau$ and $\delta g_R^\tau$, and find that
$\frac{\delta A_\tau}{A_\tau} > 0$ which is also consistent with the present
LEP experiment. These conclusions are good to the realistic one-family
ETC model$^{[7]}$  . Further improvements of the precise lepton p[olarrization
asymmetry measurements will give better constraints to the model.

  In this paper, we have compentently analyzed the corrections to the
$ Z b {\bar b} $ vertex and the $ Z \tau^+\tau^- $ vertex due to the
different possible ETC gauge exchanges in realistic one - family TC model.
The analysis for other TC models may have similar qualitative  feataures.

\end{sf}

\vspace{1cm}
\noindent {\bf ACKNOWLEDGMENT}

  This work is supported by National Natural Science Foundation of China and
the Natural Science Foundation of Henan Scientific Committee.

\vspace{4cm}
\begin{center}
{\bf Reference}
\end{center}
\begin{enumerate}

\item
 S. Weinberg, Phys. Rev. D13, 974(1976); 19, 1277(1979); L. Susskind,
 Phys. Rev. D20, 2619(1979).
\item
 M. E. Peskin and T. Tacheuchi, Phys. Rev. D46, 381(1992).
\item
 G. Altarelli and R. Barbieri, Phys. Lett. B253, 161(1991); G. Altarelli,
  R. Barbieri
and S. Jadach, Nucl. Phys. B369, 3(1992); B376, 444(E)(1994).
\item
  R. S. Chivukula, S. B. Selipsky and E. H. Simmons, Phys. Rev. Lett.
69, 575(1992); N. Evans, Phys. Lett. B331, 378(1994).
\item
 N. kitazawa, Phys. Lett. B313, 393(1993).
\item
 T. Yoshikawa, HUPD - 9415.
\item
 T. Appekquist and J. Terning, Phys. Lett. B315, 139(1993).
\item
 T. Appekquist and J. Terning, Phys. Rev. D50, 2116(1994).
\item
 R. S. Chivukula, E. Gates, E. H. Simmons and J. Terning,
 Phys. Lett. 311B, 157(1993).
\item
 Z. Xiao, L. Wan, G. Lu, J. Yang, X. Wang, L. Guo and C. Yue,
J. Phys. G20, 901(1994).
\item
 B. Jacobsen, The Top Mass from LEP $ \& $ SLC Electroweak
 Measurements,
talk presented workshop on QCD and High Energy Hadronic Interactions,
XIXth Recontres de Moriond, March, 1994.
\item
 R. S. Chivukula, E. H. Simmons and J. Terning, Phys. Lett.B331,
 383(1994).
\item
 B. Holdom, Phys. Lett. B339, 114(1994).
\item
 A. Manohar and H. Georgi, Nucl. Phys. B234, 189(1984).
\item
 H. Georgi, Weak Interactions and Modern Particle Theory, (Benjamin -
 Cummings, Menlo Park, 1984), P. 77.
\item
Guo - Hong Wu, YCTP - P16 - 94.

\end{enumerate}
\end{document}